\documentstyle[11pt]{article}
\begin{document}

\title{Comment on {\bf "Interacting holographic dark energy model and
generalized second law of thermodynamics in a non-flat universe",
\\by M.R. Setare (JCAP {\bf 01}, 023, 2007)}}
\author{K. Karami\thanks{E-mail: KKarami@uok.ac.ir}\\
\\\small{Department of Physics, University of
Kurdistan, Pasdaran St., Sanandaj, Iran}\\
}

\maketitle
\begin{abstract}
Author of Ref. \cite{Setare3}, M.R. Setare (JCAP {\bf 01}, 023,
2007), by redefining the event horizon measured from the sphere of
the horizon as the system's IR cut-off for an interacting
holographic dark energy model in a non-flat universe, showed that
the generalized second law of thermodynamics is satisfied for the
special range of the deceleration parameter. His paper includes an
erroneous calculation of the entropy of the cold dark matter. Also
there are some missing terms and some misprints in the equations of
his paper. Here we present that his conclusion is not true and the
generalized second law is violated for the present time
independently of the deceleration parameter.
\end{abstract}
\noindent{{\bf Keywords:} dark energy theory, cosmology of theories
beyond the SM}\\\\

\section{Generalized second law of thermodynamics}
\noindent Here we give the necessary corrections which should be
applied to Section 3 in Ref. \cite{Setare3}. Equations (30), (35),
(37) and (38) in Ref. \cite{Setare3} must be corrected,
respectively, as follows
\begin{equation}
{\rm
d}S_{\Lambda}=24\pi^2c^2M_{P}^2\Big(\frac{1}{3}+\omega_{\Lambda}+\frac{\Gamma}{3H}\Big)L{\rm
d}L,\label{eqSLT1}
\end{equation}

\begin{equation}
\dot{\Omega}_{\rm
m}=-\dot{\Omega}_{\Lambda}\Big(1-\frac{1}{\Omega_\Lambda}-\frac{\Omega_{k}}{\Omega_\Lambda}\Big)
+3Hr\Omega_{\Lambda}\Big[\omega_{\Lambda}+\frac{1+r}{r}\frac{\Gamma}{3H}\Big],
\end{equation}

\begin{equation}
\frac{{\rm d}S_{\rm m}}{{\rm
d}x}=\frac{8\pi^2M_{P}^2c^3}{H^2\Omega_{\Lambda}^{3/2}}\Big[[-3b^2(1+\Omega_{k})+(1+\Omega_{k}-\Omega_{\Lambda})]
\Big(\frac{c}{\sqrt{\Omega_{\Lambda}}}-\cos{y}\Big)+3rc\sqrt{\Omega_{\Lambda}}
\Big(\omega_{\Lambda}+b^2\frac{(1+r)^2}{r}\Big)\Big],\label{dSm}
\end{equation}

\begin{eqnarray}
\frac{{\rm d}S_{\rm m}}{{\rm
d}x}=\frac{8\pi^2M_{P}^2c^3}{H^2\Omega_{\Lambda}^{3/2}}
\left\{[-3b^2(1+\Omega_{k})+(1+\Omega_{k}-\Omega_{\Lambda})]
\Big(\frac{c}{\sqrt{\Omega_{\Lambda}}}-\cos{y}\Big)\right.\nonumber\\
+\frac{c}{\sqrt{\Omega_{\Lambda}}}
\Big[-(1+\Omega_{k}-\Omega_{\Lambda})\Big(1+\frac{2\sqrt{\Omega_{\Lambda}}}{c}\cos{y}+\frac{3b^2(1+\Omega_{k})}{\Omega_{\Lambda}}\Big)
\nonumber\\\left.+\frac{3b^2(1+\Omega_{k})^2}{\Omega_{\Lambda}}\Big]\right\}.
\label{dSm}
\end{eqnarray}

The geometric entropy of the horizon in Ref. \cite{Setare3} should
be corrected as $S_{\rm L}=\pi L^2/G=8\pi^2M_{P}^2L^2$ due to have a
correct dimension. Therefore, Eq. (40) in Ref. \cite{Setare3} should
be revised as
\begin{equation} \frac{{\rm d}S_{\rm L}}{{\rm
d}x}=\frac{16\pi^2M_{P}^2c}{H^2\sqrt{\Omega_{\Lambda}}}\Big(\frac{c}{\sqrt{\Omega_{\Lambda}}}-\cos{y}\Big).
\end{equation}

Finally, the corrected form of Eq. (41) in Ref. \cite{Setare3} is
given by

\begin{eqnarray}
\frac{\rm d}{{\rm d}x}(S_{\Lambda}+S_{\rm m}+S_{\rm
L})=\frac{16\pi^2M_{P}^2c^3}{H^2\sqrt{\Omega_{\Lambda}}}
\Big(\frac{\sqrt{\Omega_{\Lambda}}}{c}\cos^2{y}-\cos{y}\Big)\nonumber\\
+\frac{8\pi^2M_{P}^2c^3}{H^2\Omega_{\Lambda}^{3/2}}
\left\{[-3b^2(1+\Omega_{k})+(1+\Omega_{k}-\Omega_{\Lambda})]
\Big(\frac{c}{\sqrt{\Omega_{\Lambda}}}-\cos{y}\Big)\right.\nonumber\\
+\frac{c}{\sqrt{\Omega_{\Lambda}}}
\Big[-(1+\Omega_{k}-\Omega_{\Lambda})\Big(1+\frac{2\sqrt{\Omega_{\Lambda}}}{c}\cos{y}+\frac{3b^2(1+\Omega_{k})}{\Omega_{\Lambda}}\Big)
\nonumber\\\left.+\frac{3b^2(1+\Omega_{k})^2}{\Omega_{\Lambda}}\Big]\right\}\nonumber\\
+\frac{16\pi^2M_{P}^2c}{H^2\sqrt{\Omega_{\Lambda}}}\Big(\frac{c}{\sqrt{\Omega_{\Lambda}}}-\cos{y}\Big).\label{Stot1}
\end{eqnarray}

For $\cos y=0.99$, $\Omega_{\Lambda}=0.73$, $\Omega_{k}=0.01$,
$b^2=0.2$ and $c=1$ given by Ref. \cite{Setare3} for the present
time, we get
$$\frac{\rm d S_{\Lambda}}{{\rm d}x}=-\Big(\frac{\pi^2M_{P}^2}{H^2}\Big)2.86,$$
$$\frac{\rm d S_{\rm m}}{{\rm d}x}=-\Big(\frac{\pi^2M_{P}^2}{H^2}\Big)2.97,$$
$$\frac{\rm d S_{\rm L}}{{\rm d }x}=\Big(\frac{\pi^2M_{P}^2}{H^2}\Big)3.38.$$

Therefore the generalized second law (GSL) due to different
contributions of the holographic dark energy, cold dark matter and
horizon for the present time is obtained as
\begin{equation}
\frac{\rm d}{{\rm d}x}(S_{\Lambda}+S_{\rm m}+S_{\rm
L})=-\Big(\frac{\pi^2M_{P}^2}{H^2}\Big)2.45<0,
\end{equation}
which compared to Eq. (44) in Ref. \cite{Setare3} shows that in
contrary to the conclusion of Author of Ref. \cite{Setare3}, the GSL
is violated for the present time independently of the deceleration
parameter for the interacting holographic dark energy with cold dark
matter in a non-flat universe enveloped by the event horizon
measured from the sphere of the horizon named $L$.

\section{Conclusions}
Here we showed that the essential conclusion of Ref. \cite{Setare3}
is not true. Author of Ref. \cite{Setare3} has concluded that for an
interacting holographic dark energy with cold dark matter in the
universe enclosed by the event horizon measured from the sphere of
the horizon as the system's IR cut-off, the generalized second law
for the present time is satisfied only for the special range of the
deceleration parameter. By the help of correcting the necessary
relation for calculating the entropy of the cold dark matter and
revising some missing terms and some misprints in the equations of
Ref. \cite{Setare3}, we presented that the GSL for the present time
is violated independently of the deceleration parameter.



\end{document}